\documentclass[twocolumn,pra,groupedaddress,showpacs,amsfonts,amssymb,amsmath,letterpaper,floatfix]{revtex4-1}

\usepackage{amsmath}
\usepackage{graphicx}
\usepackage{amssymb}
\usepackage{pslatex}

\usepackage{color}

\usepackage{txfonts}
\usepackage{dcolumn}

\begin{document}

\title{Strongly correlated photons generated by coupling a three- or four-level system to a waveguide}

\author{Huaixiu Zheng}
\email{hz33@duke.edu}
\author{Daniel J. Gauthier}
\author{Harold U. Baranger}
\email{baranger@phy.duke.edu}
\affiliation{\textit{Department of Physics, Duke University, P. O. Box 90305,
Durham, North Carolina 27708, USA}}

\date{\today}

\begin{abstract}
We study the generation of strongly-correlated photons by coupling
an atom to photonic quantum fields in a one-dimensional waveguide.
Specifically, we consider a three-level or four-level system for the
atom. Photon-photon bound-states emerge as a manifestation
of the strong photon-photon correlation mediated by the atom. Effective repulsive or attractive
interaction between photons can be produced, causing either suppressed
multiphoton transmission (photon blockade) or enhanced multiphoton
transmission (photon-induced tunneling). As a result, nonclassical
light sources can be generated on demand by sending coherent states into the
proposed system. We calculate the second-order correlation
function of the transmitted field and observe bunching and anti-bunching caused by the bound-states. Furthermore, we demonstrate that the proposed system can produce photon
pairs with a high degree of spectral entanglement, which have a large
capacity for carrying information and are important for large-alphabet
quantum communication.
\end{abstract}


\maketitle

\section{Introduction}

Strong coupling between light and matter has been demonstrated both
in classical cavity quantum electrodynamics (QED) systems
\cite{MabuchiSci02, VogelQO03, ThompsonPRL92, ReithmaierNat04} and
in more recent circuit-QED experiments \cite{SchoelkopfNat08,
WallraffNat04, ChiorescuNat04, AstafievSci10}. This enables the
generation of strong nonlinear photon-photon interactions at the
single-photon level, which is of great interest for the observation
of quantum nonlinear optical phenomena \cite{ImamogluPRL97,
HarrisPRL90, Harris97, AkimovNat07}, the control of light quanta in
quantum information protocols such as quantum networking
\cite{KimbleNat08, DuanRMP10}, as well as the study of strongly
correlated quantum many-body systems using light
\cite{HartmannNatPhys06,GreentreeNatPhys06,AngPRA07,RossiniPRL07,HartmannLPR08,
ChangNatPhys08,SchmidtPRL09,CarusottoPRL09,KochPRA09,SchmidtPRL10}.
For example, both electromagnetically induced transparency (EIT)
\cite{Harris97} and photon blockade \cite{ImamogluPRL97,
GrangierPRL98, ImamogluPRL98} have been observed in recent
experiments with trapped atoms in an optical cavity
\cite{BirnbaumNat05, MuckeNat10,Tanji-SuzukiSci11} and with
superconducting qubits in a microwave resonator
\cite{AbdumalikovPRL10, LangPRL11}. Coherent transfer of quantum
states between light and stationary qubits has been demonstrated in
both cavity-QED \cite{BoozerPRL07} and circuit-QED
\cite{HofheinzNat08, HofheinzNat09} systems. In a very recent
experiment, coherent transfer of photons between three resonators
has been realized in a superconducting circuit
\cite{MariantoniNatPhys11}.

Recently, an alternative waveguide-based QED system
\cite{ChangPRL06, ShenPRL07,ShenPRA07, ChangNatPhys07, QuanPRA09,
LongoPRL10, WitthautNJP10, ZhengPRA10, RoyPRL11, KolchinPRL11,
ZhengPRL11,RephaeliPRA11,PeropadrePRA11} has emerged as a promising
candidate for achieving strong coupling between photons and atoms,
motivated by tremendous experimental progress \cite{AkimovNat07,
BajcsyPRL09, BabinecNatNanotech10, ClaudonNatPhoton10,
AstafievSci10, AbdumalikovPRL10, BleusePRL11,HoiPRL11}. The
experimental systems include a metallic nanowire coupled to a
quantum dot \cite{AkimovNat07}, cold atoms trapped inside a hollow
fiber \cite{BajcsyPRL09}, a diamond nanowire coupled to a quantum
dot \cite{BabinecNatNanotech10}, a 1D superconducting transmission
line coupled to a qubit \cite{AstafievSci10, AbdumalikovPRL10}, and
a GaAs photonic nanowire with embedded InAs quantum dots
\cite{ClaudonNatPhoton10, BleusePRL11}. In particular, it has been
experimentally demonstrated that more than $90\%$ of the
spontaneously emitted light has been guided into the desired
waveguide mode \cite{BleusePRL11}, deep into the strong-coupling
\footnote{Here, ``strong coupling'' means that the decay rate of the
excited atom to the waveguide modes dominates over the decay rate to
all other channels. It is similar but not identical to the
definition of ``strong coupling'' in the cavity case, which requires
the vacuum Rabi frequency being much larger than the atomic
spontaneous decay rate and the cavity field decay rate
\cite{MabuchiSci02}.} regime. Theoretically, single-photon switches
\cite{ChangNatPhys07, ZhouPRL08, WitthautNJP10, KolchinPRL11} have
been proposed based on a waveguide QED scheme. An interesting
photon-atom bound state and radiation trapping have been predicted
based on numerical calculations \cite{LongoPRL10, LongoPRA11}. It
has also been shown theoretically that EIT \cite{RoyPRL11,
ZhengPRL11} and photon blockade \cite{ZhengPRL11} emerge in a 1D
waveguide system.

In this work, we consider using a waveguide-QED system to generate
strongly-correlated photons through coupling to a three-level or
four-level system (3LS or 4LS). Such strongly-correlated photons can
be used to study many-body physics \cite{HartmannLPR08} as well as
to implement large-alphabet quantum communication protocols
\cite{HumblePRA07, Ali-KhanPRL07}. Specifically, to probe the strong
photon-photon correlation mediated by the 3LS or 4LS, we study
photonic transport, number statistics, second-order correlation, and
spectral entanglement of the correlated photon states. Following
Refs.\,\onlinecite{NishinoPRL09,ZhengPRL11, ZhengPRA10,RoyPRL11},
and \onlinecite{ImamuraPRB09}, we explicitly construct the
scattering eigenstates by imposing an open boundary condition and
setting the incident state to be a free plane wave. In the
multiphoton solutions, photon-photon bound-states emerge, which have
significant impact on the physical quantities described above. While
single-photon transport exhibits EIT, multiphoton transport shows
photon-induced tunneling and photon blockade. A highly entangled
photon pair in frequency is obtained by scattering a two-photon
state off the 4LS. Finally, we study the scattering of a coherent
state wavepacket, whose number statistics become non-Poissonian.
Strong bunching and anti-bunching appear in the second-order
correlation function.

This paper is organized as follows. In Sec.\,II, we introduce the model Hamiltonian, identify relevant experimental systems, and solve for the scattering eigenstates for one-, two- and three-photon states. With the scattering eigenstates, the asymptotic output states from scattering Fock states off the 3LS or 4LS are obtained in Sec.\,III. In Sec.\,IV, we study the photonic transport of Fock states and analyze the effect caused by the photon-photon bound-states. In Sec.\,V, we calculate the spectral entanglement for the two-photon case and demonstrate that highly entangled photon pairs are obtained. In Sec.\,VI, the signatures of photon correlation are revealed in the number statistics and second-order correlation function after scattering a coherent state wavepacket. Finally, we conclude in Sec.\,VII. Some results related to photon blockade in the 4LS were reported previously in Ref.\,\onlinecite{ZhengPRL11}.

\section{System, Hamiltonian, and Scattering Eigenstates}

\begin{figure}[t]
\centering
\includegraphics[width=0.5\textwidth]{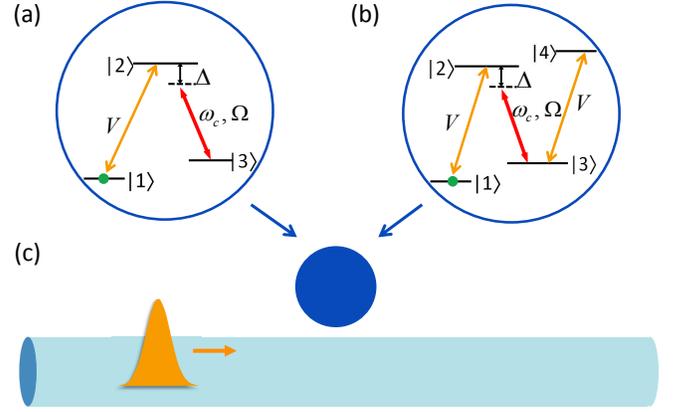}
\caption{(color online) Sketch of the atom-waveguide system:
      (a) a $\Lambda$-type three-level system,
      (b) an $N$-type four-level system,
      (c) photons (yellow) in a 1D waveguide coupled to an atom (blue), which can be either the 3LS in (a) or the 4LS in (b).
The transitions $|1\rangle\leftrightarrow|2\rangle$ and $|3\rangle\leftrightarrow|4\rangle$ are coupled to the waveguide modes with strength $V$.
The transition $|2\rangle\leftrightarrow|3\rangle$ is driven by a semiclassical control field with Rabi frequency $\Omega$ and detuning $\Delta$.
Here, $\omega_c$ is the frequency of the control field.}
\label{fig:schematic}
\end{figure}

We consider the scattering problem of photons in a one-dimensional waveguide side-coupled to a single atom, as shown in Figure\,\ref{fig:schematic}. By \textit{``atom''} we mean a local emitter with discrete levels, which could be formed from natural atoms, quantum dots, trapped ions, or superconducting qubits.

Here, two types of local emitter are considered: a driven
$\Lambda-$type 3LS and an $N$-type 4LS. The single-photon dynamics
for the 3LS was previously studied in
Ref.\,\onlinecite{WitthautNJP10} and a two-photon solution was found
in Ref.\,\onlinecite{RoyPRL11} in the limit of weak control field.
Here, without assuming a weak control field, we solve the scattering
problem for both the 3LS and 4LS in the general case. We mainly
focus on the photon-photon correlation induced by the atom:
physically, the interesting physics originates from the interplay of
quantum interference in the 1D waveguide and interaction effects
induced by the atom. Such interaction can be understood by treating
the atom as a bosonic site and the ground and excited states as zero
and one boson states, respectively. Unphysical multiple occupation
is removed by adding an infinitely large repulsive on-site
interaction term \cite{LongoPRL10}, which is the underlying
mechanism responsible for the formation of photon-photon bound
states \cite{ShenPRL07, ShenPRA07, ZhengPRA10, RoyPRL11,
ZhengPRL11}. The proposed system could be realized either in optical
systems \cite{AokiNat06,DayanSci08, AkimovNat07, ClaudonNatPhoton10,
BleusePRL11} or in microwave superconducting (SC) circuits
\cite{BianchettiPRL10, EichlerPRL11, AstafievSci10,
AbdumalikovPRL10, MajerPRL05}. For the optical systems, the driven
3LS and 4LS have been studied in both the trapped ion
\cite{SlodickaPRL10} and cavity systems \cite{MuckeNat10,
Tanji-SuzukiSci11,AlbertaNatPhotonics11, ChangPRL04}. For the
microwave SC systems, the 3LS and 4LS have already been realized
using SC qubits \cite{BaurPRL09,SillanpaaPRL09,
KellyPRL10,AbdumalikovPRL10, BianchettiPRL10, MajerPRL05}.

We start with the Hamiltonian in the rotating wave approximation,
describing a continuum photonic field in a 1D waveguide coupled
to a single atom \cite{ShenPRL07, WitthautNJP10, ZhengPRA10, RoyPRL11,
ZhengPRL11}
\begin{eqnarray}
\label{eq:Hamiltonian}
& &H =  H_{\text{wg}}+H_{\text{atom}}+H_{c},\nonumber \\
& &H_{\text{wg}}  =  \int dx(-i)\hbar c\left[a_{R}^{\dagger}(x)\frac{d}{dx}a_{R}(x)-a_{L}^{\dagger}(x)\frac{d}{dx}a_{L}(x)\right], \qquad
\end{eqnarray}
where $a_{R,L}^{\dagger}(x)$ is the creation
operator for a right- or left-going photon at position $x$ and
$c$ is the group velocity of photons.
For the driven $\Lambda-$type 3LS,
\begin{eqnarray}
&&H_{\text{atom}}^{(\Lambda)}  =  \sum_{j=2,3} \hbar\Big(\epsilon_{j}-\frac{i\Gamma_{j}}{2}\Big)|j\rangle\langle j|+\frac{\hbar\Omega}{2}\Big(|2\rangle\langle3|+{\rm h.c.}\Big), \nonumber \\
&&H_{c}^{(\Lambda)} =  \int dx\hbar V\delta(x)\Big\{ [a_{R}^{\dagger}(x)+a_{L}^{\dagger}(x)]|1\rangle\langle2|+{\rm h.c.}\Big\}. \qquad
\end{eqnarray}
For the $N$-type 4LS,
\begin{eqnarray}
&&H_{\text{atom}}^{(N)} = \sum_{j=2}^{4}\hbar\Big(\epsilon_{j}-\frac{i\Gamma_{j}}{2}\Big)|j\rangle\langle j|+\frac{\hbar\Omega}{2}\Big(|2\rangle\langle3|+{\rm h.c.}\Big), \\
&&H_{c}^{(N)} = \int dx\hbar V\delta(x)\Big\{ [a_{R}^{\dagger}(x)+a_{L}^{\dagger}(x)](|1\rangle\langle2|+|3\rangle\langle4|)+{\rm h.c.}\Big\} . \nonumber
\end{eqnarray}
Here, the energy reference is the energy of the ground state $|1\rangle$, and $\epsilon_2=\omega_{21}$,
$\epsilon_3=\epsilon_2-\Delta$, and $\epsilon_4=\epsilon_3+\omega_{43}$, where $\omega_{21}$
and $\omega_{43}$ are the $|1\rangle\leftrightarrow|2\rangle$, and $|3\rangle\leftrightarrow|4\rangle$
transition frequencies, respectively. In the spirit of the quantum
jump picture \cite{Carmichael93}, we include an imaginary
term in the energy level to model the spontaneous emission of the
excited states at rate $\Gamma_{j}$ to modes other than the waveguide
continuum.The spontaneous emission rate to the 1D waveguide continuum is given by $\Gamma=2V^{2}/c$ (from Fermi's golden rule).
Notice that the use of the rotating wave approximation is justified by the fact that $\hbar\Gamma\ll \hbar\omega_{21}$, which is the case in current experiments \cite{AstafievSci10, AbdumalikovPRL10,ClaudonNatPhoton10,BleusePRL11, HoiPRL11}.

It is convenient to transform the right/left modes to even/odd
modes: $a_{e}^{\dagger}(x) =[a_{R}^{\dagger}(x)+a_{L}^{\dagger}(-x)]
/ \sqrt{2}$ and $a_{o}^{\dagger}(x)
=[a_{R}^{\dagger}(x)-a_{L}^{\dagger}(-x)] / \sqrt{2}$. This
decomposes the Hamiltonian into two decoupled modes. The even mode
couples to the atom and the odd mode is free: $H=H_{e}+H_{o}$ with
\begin{subequations}
\begin{eqnarray}
H_{e} & = & \int dx(-i)\hbar ca_{e}^{\dagger}(x)\frac{d}{dx}a_{e}(x)+H_{\text{atom}}+H_{c}, \\
H_{o} & = & \int dx(-i)\hbar ca_{o}^{\dagger}(x)\frac{d}{dx}a_{o}(x).
\end{eqnarray}
\end{subequations}
The coupling Hamiltonian $H_c$ is now
\begin{subequations}
\begin{eqnarray}
H_{c}^{(\Lambda)} & = & \int dx\hbar\overline{V}\delta(x)\left\{ a_{e}^{\dagger}(x)|1\rangle\langle2|+{\rm h.c.}\right\} , \\
H_{c}^{(N)} & = & \int dx\hbar\overline{V}\delta(x)\left\{ a_{e}^{\dagger}(x)\left(|1\rangle\langle2|+|3\rangle\langle4|\right)+{\rm h.c.}\right\},
\end{eqnarray}
\end{subequations}
where $\overline{V}=\sqrt{2}V$.
Hereafter, we will concentrate on solving for the scattering
eigenstates in the even space. Because $[H,\,\hat{n}_{e}+\hat{n}_{\text{atom}}]=[H,\,\hat{n}_{o}]=0$
for the number operators $\hat{n}_{e/o}\equiv\int dx\;\hat{a}_{e/o}^{\dagger}(x)\hat{a}_{e/o}(x)$
and the atomic excitation $\hat{n}_{\text{atom}}$, the total number of excitations
in both the even and odd spaces are separately conserved. Therefore,
a general $n$-excitation state in the even space  ($n=n_{e}+n_{\text{atom}}$)
is given by
\begin{subequations}
\begin{eqnarray}
|\Psi_{n}^{(\Lambda)}\rangle_{e} & = & \bigg[\int dx^{n}\;g^{(n)}(x)\;\hat{a}_{e}^{\dagger}(x_{1})\cdots\hat{a}_{e}^{\dagger}(x_{n})  \\
& & +   \int dx^{n-1} \sum_{j=2,3}f_{j}^{(n)}(x)\;S_{1j}^{+} \;\hat{a}_{e}^{\dagger}(x_{1})\cdots\hat{a}_{e}^{\dagger}(x_{n-1})\bigg]|\emptyset,1\rangle, \nonumber \\
|\Psi_{n}^{(N)}\rangle_{e}  & = & \bigg[\int dx^{n}\;g^{(n)}(x)\;\hat{a}_{e}^{\dagger}(x_{1})\cdots\hat{a}_{e}^{\dagger}(x_{n})  \\
& & +    \int dx^{n-1} \sum_{j=2,3}f_{j}^{(n)}(x)\;S_{1j}^{+} \;\hat{a}_{e}^{\dagger}(x_{1})\cdots\hat{a}_{e}^{\dagger}(x_{n-1}) \nonumber \\
& & +   \int dx^{n-2}f_{4}^{(n)}(x)\;S_{14}^{+}\;\hat{a}_{e}^{\dagger}(x_{1})\cdots\hat{a}_{e}^{\dagger}(x_{n-2})\bigg]|\emptyset,1\rangle, \nonumber
\end{eqnarray}
\end{subequations}
where $|\emptyset,1\rangle$ is the zero-photon state with
the atom in the ground state $|1\rangle$ and $S_{ij}^{+}=|j\rangle\langle i|$.

The scattering eigenstates are constructed by imposing the open boundary
condition that $g^{(n)}(x)$ is a free-bosonic
plane wave in the incident region \cite{NishinoPRL09, ZhengPRA10, ZhengPRL11}. That is, for
$x_{1},\cdots,x_{n}<0$,
\begin{equation}
\label{eq:gn}
g^{(n)}(x) = \frac{1}{n!}\sum_{Q}h_{k_{1}}(x_{Q_{1}})\cdots h_{k_{n}}(x_{Q_{n}}),\qquad h_{k}(x) = \frac{e^{ikx}}{\sqrt{2\pi}},
\end{equation}
where $Q=(Q_{1},\cdots,Q_{n})$ is a permutation of $(1,\cdots,n)$.
Solving the Schr\"{o}dinger equation with this open boundary condition, we
find the scattering
eigenstates for the systems we consider here (for a detailed derivation for a two-level system,
see the Appendix of Ref.\,\onlinecite{ZhengPRA10}). Below, we present the
one-, two-, and three-photon scattering eigenstates, which have the same form for the 3LS and 4LS cases.
In the even space, the one-photon scattering eigenstate with eigenenergy $E=\hbar ck$ is given by
\begin{subequations}
\label{eq:SP_SE}
\begin{eqnarray}
&&g^{(1)}(x) \equiv  g_{k}(x)=h_{k}(x)\left[\theta(-x)+\overline{t}_{k}\theta(x)\right],\quad\quad \\
&&\overline{t}_{k} = \frac{\big[ck-\epsilon_{2}+\Delta+i\Gamma_{3}/2\big]\big[ck-\epsilon_{2}+(i\Gamma_{2}-i\Gamma)/2\big]-\Omega^{2}/4 }{\big[ck-\epsilon_{2}+\Delta+i\Gamma_{3}/2\big]\big[ck-\epsilon_{2}+(i\Gamma_{2}+i\Gamma)/2\big]-\Omega^{2}/4}, \quad\quad
\end{eqnarray}
\end{subequations}
where $\theta(x)$ is the step function. The one-photon
scattering eigenstate is exactly the same for both the 3LS and 4LS
because it takes at least two quanta to excite level $|4\rangle$: for single-photon processes, the 3LS and 4LS cases are equivalent.

For two-photon scattering, we start with a free plane
wave in the region $x_{1},x_{2}<0$, and use the Schr\"{o}dinger equation
to find the wave function first in the region $x_{1}<0<x_{2}$ and
then for $0<x_{1},x_{2}$ \cite{ZhengPRA10}. We arrive at the following
two-photon scattering eigenstate with eigenenergy $E=\hbar c(k_{1}+k_{2})$:
\begin{subequations}
\begin{eqnarray}
&&g^{(2)}(x_{1},x_{2})=\frac{1}{2!}\Big[\sum_{Q}g_{k_{1}}(x_{Q_{1}})g_{k_{2}}(x_{Q_{2}}) \\ \nonumber
       &&  \quad\quad\quad\quad\quad  + \sum_{PQ}B_{k_{P_1},k_{P_2}}^{(2)}(x_{Q_{1}},x_{Q_{2}}) \theta(x_{Q_{1}}) \Big] \;,\\
&&B_{k_{P_1},k_{P_2}}^{(2)}(x_{Q_{1}},x_{Q_{2}}) = e^{iEx_{Q_{2}}} \sum_{j=1,2} C_{j}e^{-\gamma_{j}|x_{2}-x_{1}|} \theta(x_{Q_{21}}) \;,
\end{eqnarray}
\end{subequations}
where $P=(P_{1},P_{2})$ and $Q=(Q_{1},Q_{2})$ are permutations of $(1,2)$, $\theta(x_{Q_{ij}})=\theta(x_{Q_{i}}-x_{Q_{j}})$,
and $B^{(2)}$ is a two-photon bound state---$\text{Re}[\gamma_{1,2}]>0$. Our solution applies for the general case of arbitrary strength of the control field.
Taking the weak control field limit for the 3LS case, we checked that one recovers the two-photon solution found in Ref.\,\onlinecite{RoyPRL11}.

Following the same procedure, we obtain
the three-photon scattering eigenstate with eigenenergy $E=\hbar c(k_{1}+k_{2}+k_{3})$:
\begin{widetext}
\begin{eqnarray}
\label{eq:g3}
&&g^{(3)}(x_{1},x_{2},x_{3}) =\frac{1}{3!}\Bigg\{
\sum_{Q}g_{k_{1}}(x_{Q_{1}})g_{k_{2}}(x_{Q_{2}})g_{k_{3}}(x_{Q_{3}}) +
\sum_{PQ}\Big[g_{k_{P_{1}}}(x_{Q_{1}})B_{k_{P_{2}},k_{P_{3}}}^{(2)}\!\!(x_{Q_{2}},x_{Q_{3}}) \;\theta(x_{Q_{2}}) +
B_{k_{P_{1}},k_{P_{2}},k_{P_{3}}}^{(3)}\!\!(x_{Q_{1}},x_{Q_{2}},x_{Q_{3}})\; \theta(x_{Q_{1}}) \Big] \Bigg\}, \nonumber\\
&&B_{k_{P_{1}},k_{P_{2}},k_{P_{3}}}^{(3)}(x_{Q_{1}},x_{Q_{2}},x_{Q_{3}}) = e^{i\big[k_{P_{1}}x_{Q_{2}}+(k_{P_{2}}+k_{P_{3}})x_{Q_{3}}\big]}
\Big[D_{1}\;e^{-\gamma_{1}|x_{Q_{3}}-x_{Q_{1}}|}+
D_{2}\;e^{-\gamma_{2}|x_{Q_{3}}-x_{Q_{1}}|} \nonumber \\
&&\qquad\qquad\qquad\qquad\qquad\qquad +D_{3}\;e^{-\gamma_{1}|x_{Q_{3}}-x_{Q_{2}}|-\gamma_{2}|x_{Q_{2}}-x_{Q_{1}}|}+
D_{4}\;e^{-\gamma_{2}|x_{Q_{3}}-x_{Q_{2}}|-\gamma_{1}|x_{Q_{2}}-x_{Q_{1}}|} \Big] \theta(x_{Q_{32}})\theta(x_{Q_{21}}),
\end{eqnarray}
\end{widetext}
where $B^{(3)}$ is a three-photon bound state,
$P=(P_{1},P_{2},P_{3})$ and $Q=(Q_{1},Q_{2},Q_{3})$ are permutations
of $(1,2,3)$. The coefficients $C_{1,2}$ and $D_{1,2,3,4}$ in the
bound states depend on the system parameters and have different
functional forms for the 3LS and 4LS. Expressions for
$\gamma_{1,2},\, C_{1,2}$, and $D_{1,2,3,4}$ are given in Appendix
A. Notice that the bound states here have more structure than in the
two-level case \cite{ShenPRA07, ShenPRL07, ZhengPRA10}; for example,
the two-photon bound state has {\it two} characteristic binding
strengths instead of one. This is due to the internal atomic
structure: for the 3LS or 4LS, the photonic field couples to the
transitions from the ground state to both of the eigenstates in the
dressed state picture of levels $|2\rangle$ and $|3\rangle$, giving
rise to two binding strengths. Such bound states are a manifestation
of the photon-photon correlation induced by having two or more
photons interact with the same atom. For the 4LS case, this leads to
strikingly different multiphoton transport behavior compared to the
single-photon transport \cite{ZhengPRL11}.

From the scattering eigenstates, we construct $n$-photon ($n=1$ to $3$) scattering matrices ({\it S} matrices)
using the Lippmann-Schwinger formalism \cite{SakuraiQM94, ZhengPRA10, ShenPRA07}. The output
states are then obtained by applying the {\it S} matrices on the incident states \cite{ZhengPRA10}.

\section{Output States of Fock State Scattering}

In this section, we present the output states from scattering one-, two-, and three-photon number states off of a 3LS or 4LS.
We assume that the incident state propagates to the right and the atom is initially in the ground state.
Specifically, we consider incident states in the form of a wavepacket for two reasons:
(i) in practice, any state that contains a finite number of photons is a wavepacket;
(ii) as we will show, sending in wavepackets with a finite width is crucial in order to observe the bound state effects in the measurements.
The continuous-mode photon-wavepacket creation operator is given by \cite{LoudonQTL03}
\begin{equation}
 a_{\alpha,\, R/L}^{\dagger}=\int\! dk \;\alpha(k) \;a_{R/L}^{\dagger}(k)  ,
\end{equation}
where $a_{R/L}^{\dagger}(k)=(1/\sqrt{2\pi})\int dx \;e^{ikx}a_{R/L}^{\dagger}(x)$ and the amplitude $\alpha(k)$ satisfies the normalization condition $\int dk\; |\alpha(k)|^2=1$. An incident right-going $n$-photon Fock state is defined as
\begin{equation}
 |n_{\alpha}\rangle_R=\frac{(a_{\alpha,\, R}^{\dagger})^n}{\sqrt{n!}}|\emptyset\rangle.
\end{equation}
With the $n$-photon {\it S} matrices $S^{(n)}$, we are able to find
the asymptotic output state long after the scattering ($t\rightarrow
+\infty$) \cite{ZhengPRA10}. Specifically, the single-photon output
state is given by
\begin{subequations}
\label{Output_SP}
\begin{eqnarray}
 &&|\psi^{(1)}\rangle = \int dk \alpha(k)|\phi^{(1)}(k)\rangle,\\
 &&|\phi^{(1)}(k)\rangle  =t_k |k\rangle_{R} +r_k |k\rangle_{L}, \\
 && |k\rangle_{R/L}=a_{R/L}^{\dagger}(k)|\emptyset\rangle,\\
 &&t_k \equiv (\overline{t}_k+1)/2,\,\,\,\, r_k \equiv (\overline{t}_k-1)/2.
\end{eqnarray}
\end{subequations}
The two-photon output state reads
\begin{subequations}
\label{Output_TP}
 \begin{eqnarray}
 && |\psi^{(2)}\rangle = \int dk_1 dk_2\frac{1}{\sqrt{2}} \alpha(k_1)\alpha(k_2) |\phi^{(2)}(k_1,k_2)\rangle,\quad \\
 && |\phi^{(2)}(k_1,k_2)\rangle = \int dx_1 dx_2 \Big[\frac{1}{2}t_{k_1,k_2}(x_1,x_2)a_R^{\dagger}(x_1)a_R^{\dagger}(x_2)  \quad \nonumber \\
&& \qquad + rt_{k_1,k_2}(x_1,-x_2)a_R^{\dagger}(x_1)a_L^{\dagger}(x_2)  \nonumber \\
&& \qquad +  \frac{1}{2}r_{k_1,k_2}(-x_1,-x_2)a_L^{\dagger}(x_1)a_L^{\dagger}(x_2)\Big]|\emptyset\rangle ,
\end{eqnarray}
\end{subequations}
where
\begin{eqnarray}
\label{eq:Bk1k2}
&&t_{k_1,k_2} \equiv  t_{k_1}t_{k_2}h_{k_1}(x_1)h_{k_2}(x_2)+\frac{1}{4}B^{(2)}_{k_1,k_2}(x_1,x_2)+k_1\leftrightarrow k_2,\nonumber \\
&&rt_{k_1,k_2} \equiv t_{k_1}r_{k_2}h_{k_1}(x_1)h_{k_2}(x_2)+\frac{1}{4}B^{(2)}_{k_1,k_2}(x_1,x_2)+k_1\leftrightarrow k_2, \nonumber \\
&&r_{k_1,k_2} \equiv  r_{k_1}r_{k_2}h_{k_1}(x_1)h_{k_2}(x_2)+\frac{1}{4}B^{(2)}_{k_1,k_2}(x_1,x_2)+k_1\leftrightarrow k_2, \nonumber \\
&&B^{(2)}_{k_1,k_2}(x_1,x_2)\equiv e^{i(k_1+k_2)x_2}\sum_{j=1,2}C_je^{-\gamma_j|x_2-x_1|}\theta(x_{21})+(x_1\leftrightarrow x_2).\nonumber \\
\end{eqnarray}
In Eq.\,(\ref{Output_TP}), the output state has three components
$t_{k_1,k_2}$, $rt_{k_1,k_2}$ (which is not a product), and
$r_{k_1,k_2}$, corresponding to two-photon transmission, one-photon
transmitted and one-photon reflected, and two-photon reflection,
respectively. The first term in each of these functions is the
plane-wave term. The second term is the bound-state term associated
with the momentum-nonconserved (for individual photons) processes.
The three-photon output state takes a similar form and is shown in
Appendix B.

\begin{figure}[tb]
 \centering
 \includegraphics[width=0.45\textwidth]{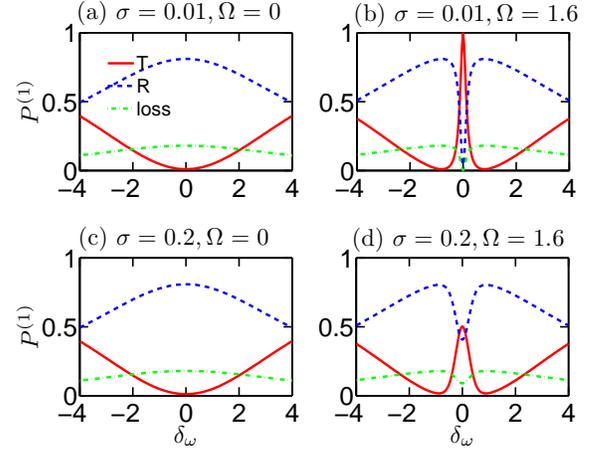}
\caption{Single-photon transmission $T$ (solid), reflection $R$
(dashed) and loss (dotted) as a function of incident photon
detuning, for the values of $\sigma$ (the wavepacket width) and
$\Omega$ (the strength of the control field) shown. Throughout the
paper, we set the loss rate of level $2$ as our frequency unit:
$\Gamma_2=1$. Here, the effective Purcell factor is $P=9$. Note the
sharp EIT window, particularly in the narrow wavepacket case. }
 \label{fig:EIT}
\end{figure}

\begin{figure*}[tb]
\centering
\includegraphics[width=0.78\textwidth]{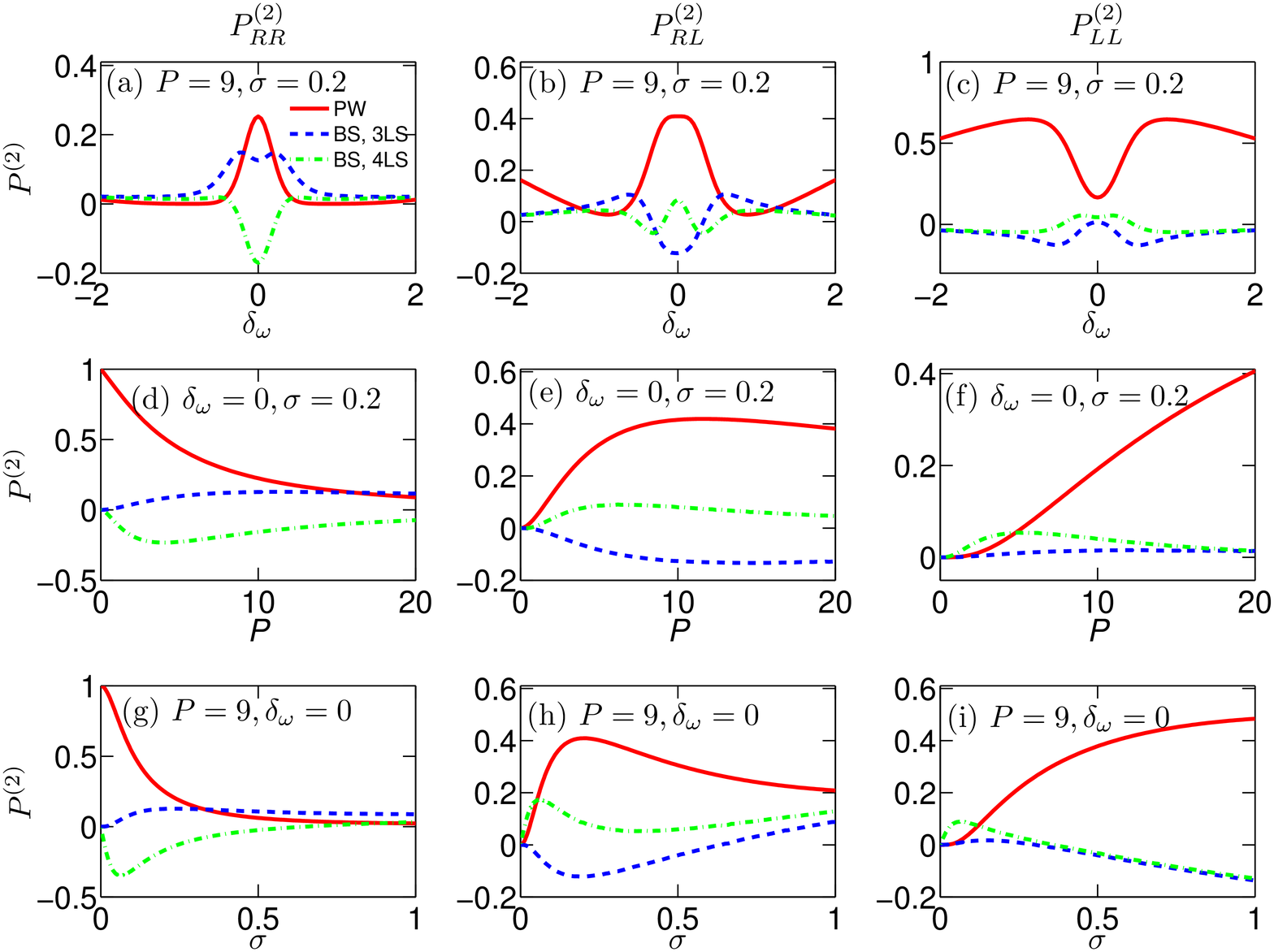}
\caption{\label{fig:TwoPhoton} (color online) Two-photon
transmission and reflection
 probabilities for the 3LS and 4LS cases.
 (a)-(c) As a function of incident photon detuning $\delta_{\omega}$ with $P=9$ and $\sigma=0.2$.
 (a) Probability that both photons are transmitted (and hence are right-going, $P^{(2)}_{RR}$).
  (b) Probability that one photon is transmitted and one reflected (right-left, $P^{(2)}_{RL}$).
  (c) Probability that both photons are reflected (both left-going, $P^{(2)}_{LL}$).
  (d)-(f) As a function of $P$ with $\delta_{\omega}=0$ and $\sigma=0.2$.
  (g)-(i) As a function of $\sigma$ with $P=9$ and $\delta_{\omega}=0$.
The label PW refers to the contribution from the plane-wave term
only, while BS refers to all the other contributions involving
bound-state terms [Eq.\,(\ref{eq:P2RR})]. Here, we set $\Omega=1.6$.
The bound state effect enhances transparency in the 3LS case but
blocks two-photon transmission past a 4LS. Note that a non-zero
$\sigma$ is crucial to observe these effects. }
\end{figure*}

With the output states, we can study induced photon-photon correlation by applying various measurements on them.
We present results for transport, spectral entanglement, number statistics, and second-order correlation in the following three sections.
Throughout the paper, we choose incident Gaussian wavepackets with the spectral amplitude
\begin{equation}
\alpha(\omega)= \frac{1}{(2\pi \sigma^2)^{1/4}}
\exp \Big[ -\frac{(\omega-\omega_0)^2}{4\sigma^2}\Big] \;,
\end{equation}
where $\sigma$ is the width and $\omega_0$ is the central frequency.
We assume that level $|3\rangle$ is metastable ($\Gamma_3$=0) and levels $|2\rangle$ and $|4\rangle$ have the same loss rate: $\Gamma_2=\Gamma_4$.
In addition, we assume that the transitions $|1\rangle\leftrightarrow|2\rangle$ and $|3\rangle\leftrightarrow|4\rangle$ are at the same frequency, $\omega_{21}=\omega_{43}$, and the detuning of the control field is zero, $\Delta=0$.
We set the loss rate as our reference frequency unit: $\Gamma_2=\Gamma_4=1$. The coupling strength to the waveguide is characterized by
the effective Purcell factor $P=\Gamma/\Gamma_2=\Gamma$.
Plasmonic waveguide systems have been predicted to have a large Purcell factor \cite{ChangNatPhys07} and a value of $P=1.5$ has been demonstrated experimentally \cite{AkimovNat07}.
Slot waveguides have been theoretically shown to have large values of $P$ reaching $16$.
Recently, by carefully tailoring the ends of photonic nanowires, J. Claudon \textit{et al.} achieved a value of $P\geq9$ in the experiment \cite{ClaudonNatPhoton10,BleusePRL11}.
Furthermore, $5.7<P<24$ was demonstrated in a photonic crystal waveguide coupled to a quantum dot \cite{Laucht12}.
In superconducting circuits with 1D open superconducting transmission lines \cite{AstafievSci10, AbdumalikovPRL10, HoiPRL11}, even larger values of $P$ have been achieved, exceeding $15$ \cite{HoiPRL11}.

\section{Transport of Few-Photon States}

\subsection{Single-Photon}

With the output state in Eq.\,(\ref{Output_SP}), the transmission ($T$),
and reflection ($R$) probabilities for a
single-photon are
\begin{subequations}
\begin{eqnarray}
 T=\int dk \,\, |_R\langle k|\psi^{(1)}\rangle|^2=\int dk \,\, \alpha^2(k)|t_k|^2, \\
 R=\int dk \,\, |_L\langle k|\psi^{(1)}\rangle|^2=\int dk \,\, \alpha^2(k)|r_k|^2,
\end{eqnarray}
\end{subequations}
which are the same for both the 3LS and 4LS cases.
Figure\,\ref{fig:EIT} shows $T$, $R$, and the loss ($1-T-R$) as a function of the detuning $\delta_{\omega}\equiv \omega_0-\omega_{21}$ at $P=9$.
Clearly, EIT appears in Fig.\,\ref{fig:EIT}(b), when the control field is on.
As one increases the width of the wavepacket, as shown in Fig.\,\ref{fig:EIT}(d), the EIT peak is suppressed as $\sigma$ becomes comparable with the width of EIT window ($\sim \Omega^2/\Gamma$), see Eqs.\,(\ref{eq:SP_SE}) and (\ref{Output_SP}).
In Fig.\,\ref{fig:EIT}(a), and (c), we set $\Omega=0$, which means the control field is off and the 3LS (4LS) becomes a reflective two-level system \cite{ChangNatPhys07, ZhengPRA10, ZhengPRL11}.
Notice that the width of the reflective peak in the $\Omega=0$ case is $\sim\Gamma$ and hence is insensitive to the increase of $\sigma$ from $0.01$ to $0.2$.

\subsection{Two-Photon}
The two-photon transmission and reflection probabilities are given by
\begin{subequations}
 \begin{eqnarray}
  P^{(2)}_{RR}&=&\int dk_1 dk_2 \frac{1}{2}|_{RR}\langle k_1,k_2|\psi^{(2)}\rangle|^2, \\
  P^{(2)}_{RL}&=&\int dk_1 dk_2 |_{RL}\langle k_1,k_2|\psi^{(2)}\rangle|^2, \\
  P^{(2)}_{LL}&=&\int dk_1 dk_2 \frac{1}{2}|_{LL}\langle k_1,k_2|\psi^{(2)}\rangle|^2,
 \end{eqnarray}
\end{subequations}
where $P^{(2)}_{RR}$, $P^{(2)}_{RL}$, and $P^{(2)}_{LL}$ are the
probabilities to observe two transmitted photons, one transmitted
and one reflected photons, and two reflected photons, respectively.
We separate the two-photon transmission and reflection probabilities
into two parts: $(P^{(2)})_\text{PW}$ is the contribution from
independent single-particle transmission (denoted PW for ``plane
wave''), and $(P^{(2)})_\text{BS}$ is the contribution from both the
bound-state term in Eq.\,(\ref{Output_TP}) and the interference
between the plane wave and bound-state terms. As an example,
$P^{(2)}_{RR}$ is split as follows
\begin{subequations}
 \begin{eqnarray}
P^{(2)}_{RR}&=&\int dk_1dk_2|\tilde{t}_2(k_1,k_2)+\tilde{B}(k_1,k_2)|^2  \nonumber \\
           &=&(P^{(2)}_{RR})^{\;}_{\text{PW}}+(P^{(2)}_{RR})^{\;}_{\text{BS}},  \\[6pt]
(P^{(2)}_{RR})^{\;}_{\text{PW}}&=&\int dk_1dk_2|\tilde{t}_2(k_1,k_2)|^2,  \\[6pt]
(P^{(2)}_{RR})^{\;}_{\text{BS}}&=&\int dk_1dk_2\big[\tilde{t}_2^*(k_1,k_2)\tilde{B}(k_1,k_2) \nonumber \\
&&+\tilde{t}_2(k_1,k_2)\tilde{B}^*(k_1,k_2)+|\tilde{B}(k_1,k_2)|^2\big],
 \end{eqnarray}
\label{eq:P2RR}
\end{subequations}
where
\begin{eqnarray}
\label{eq:P2RR_BS}
 \tilde{t}_2(k_1,k_2)&=&\alpha(k_1)\alpha(k_2)t_{k_1} t_{k_2}, \nonumber \\[6pt]
\tilde{B}(k_1,k_2)&=&\frac{i}{4c}\sum_{j=1,2}\Big(\frac{1}{k_1+i\gamma_j}+\frac{1}{k_2+i\gamma_j}\Big) \\
          && \times \int dk \;\alpha(k)\;\alpha(k_1+k_2-k)\;C_j(k,k_1+k_2-k) \;. \nonumber
\end{eqnarray}

Figure\,\ref{fig:TwoPhoton} shows the two-photon transmission and reflection probabilities for both the 3LS and 4LS cases, decomposed in this way.
Because the PW term is from the single-particle solution, it is the same for both the 3LS and 4LS.
However, $(P^{(2)})_{\text{BS}}$ is quite different for the 3LS and 4LS. Figure\,\ref{fig:TwoPhoton}(a)-(c) shows $P^{(2)}$ as a function
of incident photon detuning.
Close to resonance, in the 3LS case $(P^{(2)})_\text{BS}$ enhances the two-photon transmission $P^{(2)}_{RR}$ while suppressing $P^{(2)}_{RL}$.
In contrast, in the 4LS case $(P^{(2)})_\text{BS}$ has exactly the opposite effect.
\emph{This leads to enhanced multiphoton EIT for the 3LS \cite{RoyPRL11} and photon blockade for the 4LS \cite{ZhengPRL11}.}
Such enhanced EIT and photon blockade are caused by the interference between the two multiphoton scattering pathways:
passing by the atom as independent particles or a composite particle in the form of bound states (for a detailed analysis, see the Supplementary Material of Ref.\,\onlinecite{ZhengPRL11}).

In Fig.\,\ref{fig:TwoPhoton}(d)-(f), we plot $P^{(2)}$ as a function the effective Purcell factor $P$ for the on-resonance case, $\delta_{\omega}=0$.
It is remarkable that, for $P^{(2)}_{RR}$ and $P^{(2)}_{RL}$, $(P^{(2)})_\text{BS}$ becomes comparable to $(P^{(2)})_\text{PW}$ in the strong coupling regime.
An important implication is that the bound-state effect can be observed in photonic transport experiments, given recent rapid experimental advances \cite{AstafievSci10, AbdumalikovPRL10,ClaudonNatPhoton10,BleusePRL11, HoiPRL11}.

Fig.\,\ref{fig:TwoPhoton}(g)-(i) shows $P^{(2)}$ as a function of the wavepacket width $\sigma$ with $P=9$ and the photons on resonance with the atom. There are several notable features. First, as $\sigma$ approaches zero, $(P^{(2)})_\text{BS}$ shrinks to zero for both the 3LS and 4LS cases.
This further highlights that sending in a wavepacket with a finite width is crucial to observe the bound state effect in photonic transport.
Physically, this occurs because, in the $\sigma=0$ limit under EIT conditions,
the atom is fully transparent ($T=1$) to the incoming photons and hence the atom-mediated photon-photon interaction is absent, inhibiting any bound state effect.
For the general case without EIT conditions, the above conclusion still holds: as $\sigma\rightarrow0$, the bound state effect vanishes in multiphoton transport.
This is because the bound-state term in Eq.\,(\ref{eq:P2RR_BS}) originates from the coincident photons at the atomic site: as $\sigma\rightarrow0$, the wavepacket becomes infinitely long and the probability of coincidence vanishes.
Second, notice that while $(P^{(2)})_\text{BS}$ approaches zero for $P^{(2)}_{RR}$ as $\sigma$ increases, its magnitude for $P^{(2)}_{RL}$ and $P^{(2)}_{LL}$ increases after an initial decrease.
This is due to the enhanced interference between the plane-wave and bound-state terms [Eq.\,(\ref{Output_TP})] for $P^{(2)}_{RL}$ and $P^{(2)}_{LL}$.

The result for three photon scattering shows behavior similar to the two-photon case. To avoid duplication, we do not present it here.

 \begin{figure}[tb]
 \centering
 \includegraphics[width=0.48\textwidth]{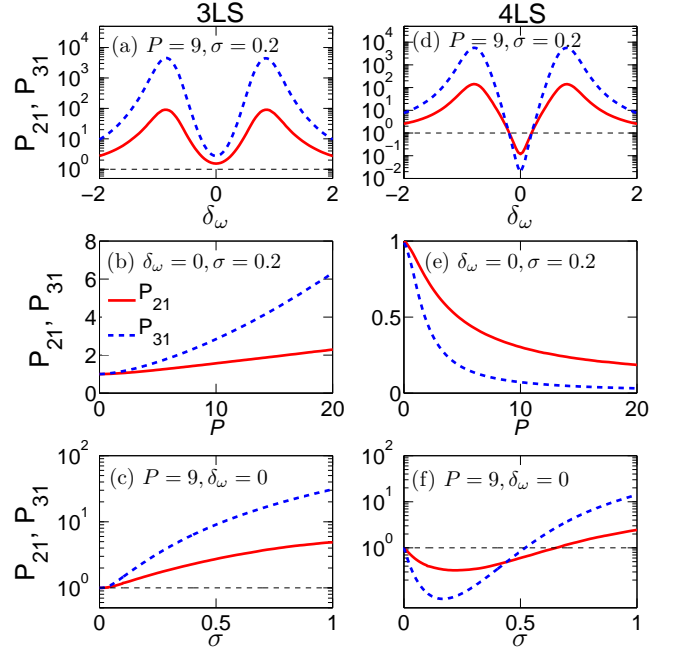}
\caption{Photon blockade and photon-induced tunneling in transmission.
Photon blockade strengths $P_{21}$ (solid line) and $P_{31}$ (dashed line) as a function of incident photon detuning $\delta_{\omega}$, $P$ and $\sigma$ for
      (a)-(c) the 3LS case,  and (d)-(f) the 4LS case.
      Here, $\Omega=1.6$. The 3LS causes photon-induced tunneling while the 4LS causes photon blockade.
}
 \label{fig:PB_PIT}
 \end{figure}

\begin{figure*}[tb]
\centering
\includegraphics[width=0.84\textwidth]{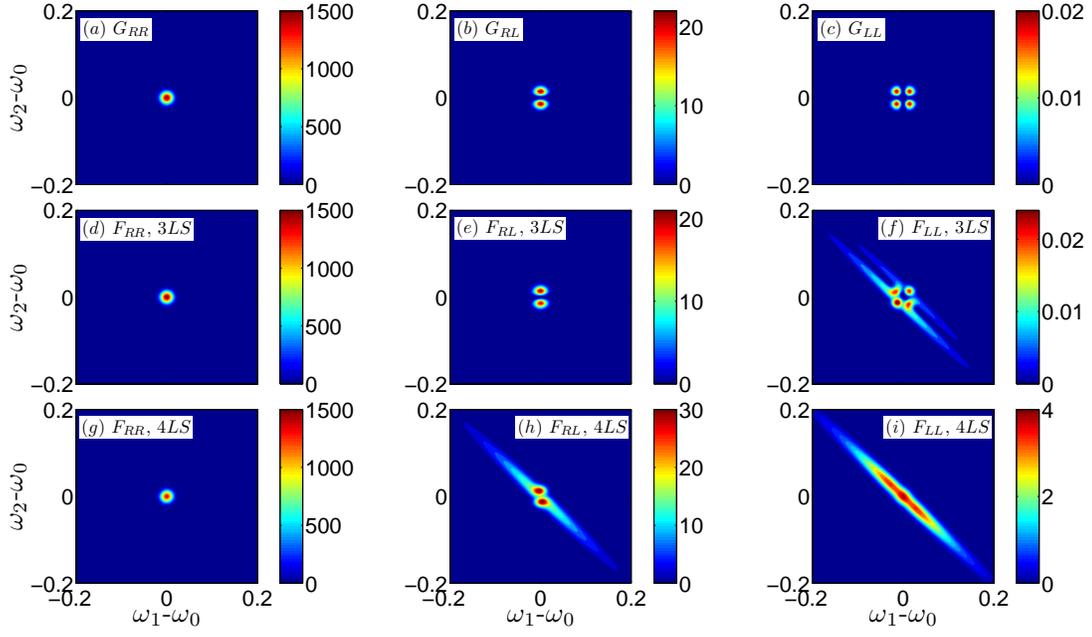}
\caption{Two-photon joint spectrum of the output states after
scattering off a 3LS and 4LS in the case of a spectrally narrow
incident wavepacket. Panels (a)-(c) show the uncorrelated spectra
$G_{RR}(\omega_1,\omega_2)$, $G_{RL}(\omega_1,\omega_2)$, and
$G_{LL}(\omega_1,\omega_2)$, respectively. Panels (d)-(f) show the
joint spectra of the transmitted field after scattering off a 3LS
for two transmitted photons, for one transmitted and one reflected,
and for two reflected photons, respectively. Panels (g)-(i) show the
joint spectra of the transmitted field after scattering off a 4LS.
Strong spectral entanglement is indicated in panel (i); this
reflected field is essentially a pure two-photon bound state. System
parameters: $P=9$, $\Omega=1.6$, $\delta_{\omega}=0$, $\sigma=0.01$.
Throughout the paper, we set the loss rate of level 2 as our
frequency unit: $\Gamma_2=1$.}
\label{fig:SE}
\end{figure*}

\subsection{Photon Blockade and Photon-Induced Tunneling}

To quantify the observed enhancement of EIT and photon blockade in Fig.\,\ref{fig:TwoPhoton}, we define the strength of photon blockade $P_{21}$ for the two-photon case by the conditional probability for transmitting a second photon given that the first photon has already been transmitted, normalized by the single-photon transmission probability.
Similarly, we can define $P_{31}$ for the three-photon case. We thus have
\begin{equation}
P_{21}\equiv \frac{P^{(2)}_{RR}}{T^2},\qquad P_{31}\equiv \frac{P^{(3)}_{RRR}}{T^3} \;.
\end{equation}
As shown in Fig.\,\ref{fig:PB_PIT}(a)-(c), for the 3LS case, the single-photon EIT is enhanced in two-photon and three-photon transmission by interaction with the 3LS.
Pronounced photon-induced tunneling \cite{FaraonNatPhy08} due to the strong correlations between transmitted photons occurs in this case: $P_{21}, P_{31}>1$.
In contrast, as shown in Fig.\,\ref{fig:PB_PIT}(d) and (e),
scattering from a 4LS exhibits a different behavior within the EIT window, namely, photon blockade \cite{ZhengPRL11}: $P_{21}, P_{31}<1$.
For increasing coupling strength [Fig.\,\ref{fig:PB_PIT}(e)], $P_{21}$ and $P_{31}$ approach zero asymptotically when the incident photons are on resonance with the 4LS.
In addition, from Fig.\,\ref{fig:PB_PIT}(c) and (f), we confirm that both photon blockade and photon-induced tunneling go away in the zero-width limit ($\sigma\rightarrow 0$).

 \begin{figure*}[tb]
 \centering
 \includegraphics[width=0.95\textwidth]{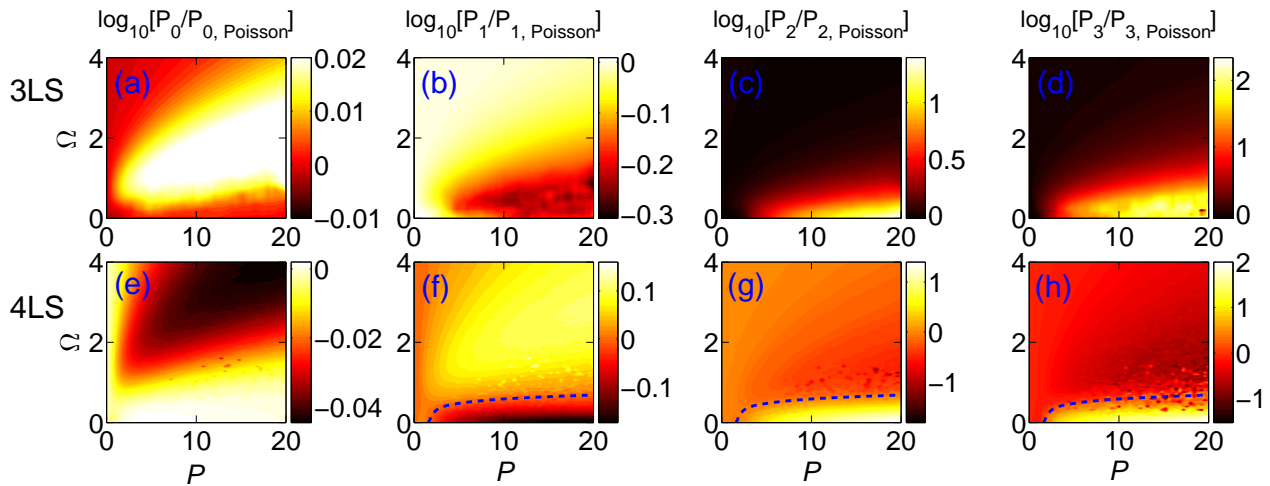}
\caption{Nonclassical light source.
Photon number statistics quantified by $\text{log}_{10}(P_n/P_{n,\text{Poisson}})$, where $P_n$ and $P_{n,Poisson}$ are the $n$-photon probability in the transmitted field and in a coherent
state with the same mean photon number, respectively.
Panels (a)-(d) show the results of the transmitted field after scattering off the 3LS for $n=0,1,2,3$, respectively.
Panels (e)-(h) show the results of the transmitted field after scattering off the 4LS for $n=0,1,2,3$, respectively.
The dashed line is a guide to the eye for equal probabilities, $P_n/P_{n,\text{Poisson}}=1$.
The speckle in the plots is numerical noise, coming from numerical evaluation of high-dimensional integrals in computing the transmission and reflection probabilities.
System parameters: $P=9$, $\Omega=1.6$, $\delta_{\omega}=0$, $\sigma=0.01$, and $\overline{n}=1$ in the incident coherent state.
Scattering off a 3LS enhances the multiphoton content of the pulse because of multi-photon EIT; in contrast, the photon blockade in the 4LS case suppresses essentially all multi-photon content, thus realizing a single-photon source.
}
 \label{fig:NS}
 \end{figure*}

\section{Spectral Entanglement of Photon Pairs}
It is clear that the two-photon bound state in Eq.\,(\ref{Output_TP}) is entangled in the momentum (or equivalently frequency) degree of freedom.
To probe this spectral aspect of the two-photon entanglement, we rewrite the two-photon output state [Eq.\,(\ref{Output_TP})] in frequency space as
\begin{eqnarray}
 |\psi^{(2)}\rangle &=&\int d\omega_1 d\omega_2 \Big[ f_{RR}(\omega_1,\omega_2) a^{\dagger}_R(\omega_1)a^{\dagger}_R(\omega_2)\\ \nonumber
                           &&\qquad + f_{RL}(\omega_1,\omega_2) a^{\dagger}_R(\omega_1)a^{\dagger}_L(\omega_2) \\ \nonumber
                           &&\qquad + f_{LL}(\omega_1,\omega_2) a^{\dagger}_L(\omega_1)a^{\dagger}_L(\omega_2) \Big]|\emptyset\rangle,
\end{eqnarray}
where $f_{RR}(\omega_1,\omega_2)$, $f_{RL}(\omega_1,\omega_2)$, and $f_{LL}(\omega_1,\omega_2)$ are the two-photon amplitudes for a transmitted pair, a pair of one transmitted and one reflected, and a reflected pair, respectively.
Explicitly, they take the following form
\begin{subequations}
\label{eq:f_RR_RL_LL}
 \begin{eqnarray}
  f_{RR}(\omega_1,\omega_2) &=& \tilde{t}_2(\omega_1,\omega_2)+\tilde{B}(\omega_1,\omega_2) ,\\
  f_{RL}(\omega_1,\omega_2) &=& 2[\tilde{rt}(\omega_1,\omega_2)+\tilde{B}(\omega_1,\omega_2)] ,\\
  f_{LL}(\omega_1,\omega_2) &=& \tilde{r}_2(\omega_1,\omega_2)+\tilde{B}(\omega_1,\omega_2) ,\\
  \tilde{t}_2(\omega_1,\omega_2)    &=& t_{\omega_1}t_{\omega_2}\alpha(\omega_1)\alpha(\omega_2), \\
  \tilde{rt}(\omega_1,\omega_2)     &=& t_{\omega_1}r_{\omega_2}\alpha(\omega_1)\alpha(\omega_2), \\
  \tilde{r}_2(\omega_1,\omega_2)    &=& r_{\omega_1}r_{\omega_2}\alpha(\omega_1)\alpha(\omega_2),
 \end{eqnarray}
\end{subequations}
where $\tilde{B}(\omega_1,\omega_2)$ is given in
Eq.\,(\ref{eq:P2RR_BS}). The first term in $f(\omega_1,\omega_2)$ is
the uncorrelated contribution, while the second term signals photon
correlation. From Eq.\,(\ref{eq:f_RR_RL_LL}), we define the joint
spectral function of the two-photon states to be \cite{BaekPRA08}
 \begin{eqnarray}
  F_{\alpha \beta=RR,\,RL,\,LL}(\omega_1,\omega_2) &=& |f_{\alpha\beta}(\omega_1,\omega_2)|^2 \;.
 \end{eqnarray}
For the purpose of comparison, we also define the uncorrelated spectral function of the two-photon states,
\begin{subequations}
 \begin{eqnarray}
  G_{RR}(\omega_1,\omega_2) &\equiv& |\tilde{t}_2(\omega_1,\omega_2)|^2, \\
  G_{RL}(\omega_1,\omega_2)  &\equiv& 4|\tilde{rt}(\omega_1,\omega_2)|^2, \\
  G_{LL}(\omega_1,\omega_2) &\equiv& |\tilde{r}_2(\omega_1,\omega_2)|^2 \;.
 \end{eqnarray}
\end{subequations}

Figure \ref{fig:SE} shows the two-photon uncorrelated and joint
spectra in the case of on-resonance photons ($\delta_{\omega}=0$)
and for a spectrally narrow wavepacket ($\sigma=0.01$). With the
chosen parameters, the EIT peak width is much larger than the
wavepacket, $\sim\Omega^2/\Gamma\cong0.28\gg\sigma$. Therefore, for
the uncorrelated pair of transmitted photons [$G_{RR}$,
Fig.\,\ref{fig:SE}(a)], there is only a sharp peak at
$\omega_1=\omega_2=\omega_0$ caused by the Gaussian spectrum of the
incident photons. For the uncorrelated pair of one transmitted and
one reflected photons ($G_{RL}$), there are two peaks resulting from
the interplay of the spectrum of the incident photons and the rapid
increase of the reflection probability away from the EIT peak (see
Fig.\,\ref{fig:EIT}). Accordingly, there are four peaks for the case
of two reflected photons, as shown in Fig.\,\ref{fig:SE}(c).

Figure\,\ref{fig:SE}(d)-(f) shows the joint spectra for the case of 3LS scattering.
It is evident that the joint spectra of the pair of two transmitted photons ($F_{RR})$, and the pair of one transmitted and one reflected photons ($F_{RL}$), are dominated by the uncorrelated
transmission. The joint spectrum of the pair of two reflected photons [Fig.\,\ref{fig:SE}(f)] is slightly modified from the uncorrelated spectrum along the diagonal line.
This is caused by the correlated bound state term $\tilde{B}(\omega_1,\omega_2)$. For the 3LS case with the chosen parameters, the correlation term $\tilde{B}(\omega_1,\omega_2)$ is of order $10^{-1}$ and hence is too weak to affect $F_{RR}$ and $F_{RL}$.

In contrast, for the 4LS case [Fig.\,\ref{fig:SE}(g)-(i)], $F_{RL}$ and $F_{LL}$ are greatly modified
by the correlation term, while $F_{RR}$ is still dominated by the uncorrelated transmission.
In particular, as shown in Fig.\,\ref{fig:SE} (i), the joint spectrum of the reflected pair is dominated by $\tilde{B}(\omega_1,\omega_2)$.
\emph{This pair is primarily made up of a pure two-photon bound state:} the frequencies of the photon pair are correlated along the line $\omega_1+\omega_2=2\omega_0$ with uncertainty $\sigma$.
A similar correlated photon pair was obtained in a waveguide-cavity system \cite{LiaoPRA10}.

The two-photon bound state is a composite object of photons with effective \emph{attractive} interaction; it displays strong bunching behavior in photon-photon correlation measurements.
Such a photon pair is highly entangled in frequency because measurement of the frequency of one photon unambiguously determines that of the other.
This strong spectral correlation provides more information per photon pair and could be used to implement large-alphabet quantum communication \cite{Ali-KhanPRL07}.

 \begin{figure}[tb]
 \centering
 \includegraphics[width=0.48\textwidth]{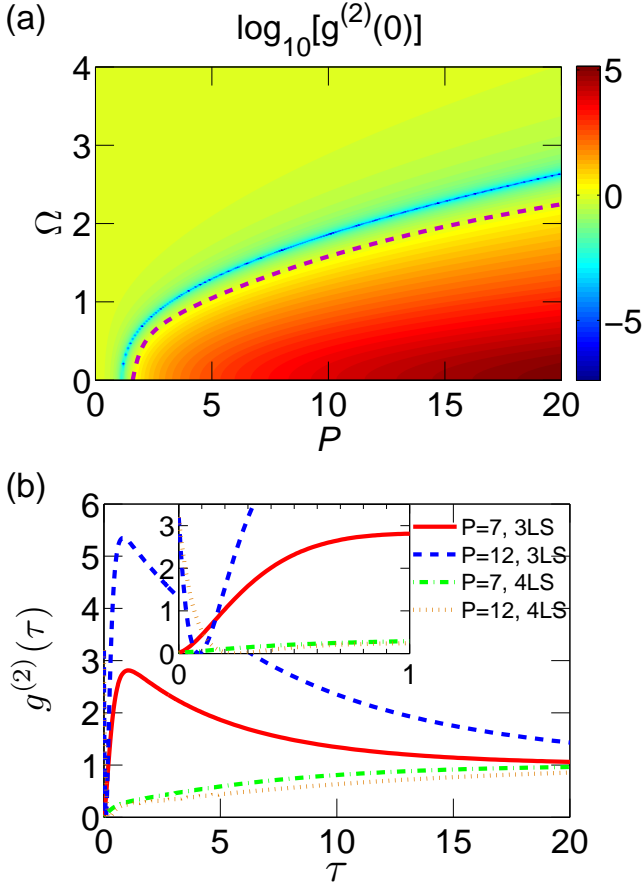}
\caption{Bunching and anti-bunching. Second-order correlation
function $g^{(2)}(\tau)$ of the transmitted field for a weak
incident coherent state ($\overline{n}\ll 1$) of width $\sigma=0.2$,
resonant with the atom ($\delta_{\omega}=0$). (a) Color map plot of
$\text{log}_{10}[g^{(2)}(0)]$ as a function of the strength of the
classical control field, $\Omega$, and the effective Purcell factor
$P$. The dashed line marks the border between bunching
($g^{(2)}(0)>1$) and anti-bunching ($g^{(2)}(0)<1$) behavior. (b)
$g^{(2)}$ as a function of time delay $\tau$ in four cases (using
$\Omega=1.6$). $\tau$ is in units of $\Gamma_2^{-1}$. Inset: zoom at
short time scales. }
 \label{fig:g20}
 \end{figure}

\section{Coherent-State Scattering}
In this section, we study the scattering of a coherent state off a 3LS or 4LS.
We probe the strong photon-photon correlation in the transmitted field by studying first the number statistics and then the second-order correlation function.

\subsection{Number Statistics}
We consider the case that the 3LS or 4LS is in its ground state initially and there is an incident continuous-mode coherent state of mean photon number $\overline{n}=1$,
spectral width $\sigma=0.2$, and central frequency on resonance with the atom, $\omega_0=\omega_{21}$.
In this case, the contribution from the four-photon state can be neglected ($\sim1.6\%$).
The photon-number statistics in the transmitted field is obtained by first applying the {\it S} matrices to the incident state and then measuring the
transmitted field, as described in Ref.\,\onlinecite{ZhengPRA10}.

We present the results for both the 3LS and 4LS cases in Fig.\,\ref{fig:NS} by taking the ratio of the photon-number distribution in the transmitted field $P_n$ ($n=0,1,2,3$) to that of a coherent state $P_{n,\text{Poisson}}$ having the same mean photon number as the transmitted field.
From Fig.\,\ref{fig:NS}(a)-(d), it is clear that when the EIT condition is satisfied, the 3LS induces strong photon-photon interactions, which in turn reduce the one-photon probability and redistributes the weight to the two- and three-photon probabilities.
This comes about because the bound state in the 3LS case enhances multiphoton EIT, as we have shown in Sec.\,IV B and C.

In contrast, for the 4LS case shown in Fig.\,\ref{fig:NS}(e)-(h), in most of the parameter space, we have enhanced single-photon probability while suppressed multiphoton content: $P_1>P_{1,\text{Poisson}}$ and $P_{2(3)}<P_{2(3),\text{Poisson}}$.
This gives rise to a sub-Poissonian single-photon source \cite{ZhengPRL11}, which comes about because, while EIT occurs in the single-photon transmission, multiphoton states experience photon blockade, as shown in Sec.\,IV B and C.
Therefore, we demonstrate that the waveguide-atom system is capable of generating nonclassical light, which may find applications in quantum cryptography \cite{LoPRL05, WangPRL05, RosenbergPRL07, GisinNatPhoton07} or distributed quantum networking \cite{KimbleNat08,DuanRMP10}.

\subsection{Second-order Correlation}

To further probe the nonclassical character of the transmitted field, we calculate the second-order correlation function $g^{(2)}(\tau)$, which is often measured experimentally.
For a steady state, $g^{(2)}$ of the transmitted field is defined as
\begin{equation}
\label{eq:g2_H}
 g^{(2)}(\tau)=\underset{t\rightarrow \infty}{\text{lim}} \frac{ \langle a^{\dagger}_R(x,t)\;a^{\dagger}_R(x,t+\tau)\; a_R(x,t+\tau)\; a_R(x,t) \rangle }{ \langle a^{\dagger}_R(x,t) \; a_R(x,t) \rangle \langle a^{\dagger}_R(x,t+\tau)\; a_R(x,t+\tau) \rangle }.
\end{equation}
As shown in Appendix C, for our system, this definition is equivalent to following expression in the Schr\"{o}dinger picture,
\begin{equation}
\label{eq:g2_S}
 g^{(2)}(\tau)=\frac{ \langle \psi | a^{\dagger}_R(x)\; a^{\dagger}_R(x+c\tau)\; a_R(x+c\tau)\; a_R(x) |\psi \rangle }{ \langle  \psi | a^{\dagger}_R(x)\; a_R(x) |\psi \rangle \langle \psi| a^{\dagger}_R(x+c\tau) \; a_R(x+c\tau) |\psi \rangle },
\end{equation}
where $|\psi\rangle$ is the asymptotic output state.
With a weak
incident coherent state (mean photon number $\overline{n}\ll1$), we consider
only the contribution of the two-photon and one-photon states in
the numerator and denominator in Eq.\,(\ref{eq:g2_S}), respectively.
Substitution of the single-photon and two-photon transmission wavefunctions from Eqs.\,(\ref{Output_SP}) and (\ref{Output_TP})
into Eq.\,(\ref{eq:g2_S}) yields the explicit expression
\begin{eqnarray}
 g^{(2)}(\tau)&=&\frac{|\int dk_1dk_2 \; \alpha(k_1) \; \alpha(k_2)\; [t_{k_1}t_{k_2}(e^{-ik_1\tau}+e^{-ik_2\tau})+B(\tau)]|^2}  {|\int dk_1dk_2\; \alpha(k_1)\; \alpha(k_2) \; t_{k_1}t_{k_2}(e^{-ik_1\tau}+e^{-ik_2\tau})|^2},\nonumber  \\
 B(\tau)&=&\pi( C_1 e^{-\gamma_1c\tau}+C_2 e^{-\gamma_2c\tau} ) \;.
\label{eq:g2}
\end{eqnarray}
In the numerator, the first term and the second term $B(\tau)$ come from the plane wave and bound state pieces, respectively, in Eq.\,(\ref{Output_TP}).

Figure\,\ref{fig:g20}(a) shows $g^{(2)}(0)$, which is the same for
the 3LS and 4LS cases. The presence of level $|4\rangle$ does not
contribute to $g^{(2)}(0)$: it takes two quanta to excite
$|4\rangle$, which then undergoes cascaded emission with zero
probability to emit two photons at the same time. In
Fig.\,\ref{fig:g20}(a), there is rich bunching and anti-bunching
behavior, caused by the two-body bound state. At $\tau$=0, the
amplitude of the bound state term in Eq.\,(\ref{eq:g2}) is
$B(0)=-2r_{k_1}r_{k_2}$, where $r_{k_{1(2)}}$ is the single-photon
reflection coefficient. Hence, in the numerator of $g^{(2)}(0)$, the
amplitudes of the plane-wave and bound-state terms are out of phase.
When $P=0$, the bound state term is zero and $g^{(2)}(0)=1$. As $P$
increases, the strength of the bound state increases, causing
$g^{(2)}(0)$ to decrease until the bound state term cancels the
plane wave term exactly, producing complete anti-bunching. Further
increase of $P$ leads to a rise of $g^{(2)}(0)$ and eventually
photon bunching.

By comparing Fig.\,\ref{fig:PB_PIT}(c)-(d) and
Fig.\,\ref{fig:g20}(a), we find that photon anti-bunching and photon
blockade, and photon bunching and photon-induced tunneling do {\it
not} have a one-to-one correspondence. For example, in the whole
parameter regime of of Fig.\,\ref{fig:PB_PIT}(d), photon blockade is
present; while in Fig.\,\ref{fig:g20}, there is a large region of
parameter space where photon bunching [$g^{(2)}(0)>1$] instead of
photon anti-bunching [$g^{(2)}(0)<1$] is observed. This is because
we are studying a state of continuous modes and performing
instantaneous measurements at two space-time points ($x, t$) and
($x, t+\tau$). If one integrates over the time $t$ in the
measurement \cite{LoudonQTL03}, as done in many experiments in which
the detector integration time is much longer than the wavepacket
duration, one finds a one-to-one correspondence between photon
anti-bunching and photon blockade, and photon bunching and
photon-induced tunneling.

The time dependence of $g^{(2)}(\tau)$ is shown in Fig.\,\ref{fig:g20}(b). There are two characteristic time scales: $\tau_1=1/\text{Re}[c\gamma_1]$ and $\tau_2=1/\text{Re}[c\gamma_2]$.
Within the short time scale, $g^{(2)}$ can display either bunching or anti-bunching for both the 3LS and 4LS cases, depending on the system parameters, as shown in the inset of Fig.\,\ref{fig:g20}.
On the long time scale, for the 3LS case, $g^{(2)}$ shows bunching---$g^{(2)}(\tau)>1$---corresponding to the enhanced multiphoton transmission already apparent from both the photon-induced tunneling [Fig.\,\ref{fig:PB_PIT}(c)] and the enhanced multiphoton content in the number statistics [Fig.\,\ref{fig:NS}].
For the 4LS case, anti-bunching [$g^{(2)}(\tau)<1$] dominates at long times, corresponding to the photon blockade observed in Fig.\,\ref{fig:PB_PIT}(d) and the enhanced single-photon content in Fig.\,\ref{fig:NS}.
{\it Hence, for our pulsed output state, $g^{(2)}(\tau=0)$ displays rich physics due to the induced photon-photon correlation, but is not necessarily a good guide to the photon statistics.}

\section{Conclusions}
In summary, we present a waveguide-QED-based scheme to generate strongly-correlated photons, of interest for both many-body physics and quantum information science.
Photon bound-states appear in the scattering eigenstates as a manifestation of the photon-photon correlation. As a result, while a single-photon experiences EIT in the proposed waveguide-atom system, multiphoton states can display either photon blockade or photon-induced tunneling, depending on the detailed structure of the ``atom''. From either the photon blockade or photon-induced tunneling that occurs, nonclassical light sources can be generated by sending coherent states into the system.
In the most interesting case, a 4LS removes the multiphoton content from the coherent state, leaving a pulse with only zero or single photon content.

In addition, we find that the system can be used to produce highly
entangled photon pair states in frequency space, potentially of use
for large alphabet quantum communication. Finally, we show that rich
bunching or anti-bunching behavior is present in the second-order
correlation function as a signature of the strong photon-photon
correlation mediated by the ``atom''. Given the recent rapid
experimental advances in several realizations, the proposed
waveguide-QED system is emerging as a promising route to {\it
cavity-free open quantum networks}, which are crucial for both
large-scale quantum computation and long-distance quantum
communication.

\section*{Acknowledgments}
We would like to thank P. G. Kwiat and W. P. Grice for calling our
attention to the importance of spectral entanglement. This work was
supported by the U.S.\,NSF Grant No. PHY-10-68698. H.Z.\ is
supported by the John T.\ Chambers Fellowship from the Fitzpatrick
Institute for Photonics of Duke University.

\baselineskip13pt
\renewcommand{\theequation}{A\arabic{equation}}
\setcounter{equation}{0}  
\begin{widetext}

\section*{Appendix A: Expressions for $\gamma_{1,\,2}$, $C_{1,\,2}$ and $D_{1,\,2,\,3,\,4}$}
In this Appendix, we give explicit expressions for the constants $\gamma_{1,\,2}$, $C_{1,\,2}$, and $D_{1,\,2,\,3,\,4}$ that appear in Eqs.\,(9) and (10) for both the 3LS and 4LS scattering eigenstates.
$\gamma_{1,2}$ is the same for both cases and is given by
\begin{subequations}
\begin{eqnarray}
 && c\gamma_{1} =  \frac{\Gamma+\Gamma_{2}+\Gamma_{3}}{4}-\xi
         +i\bigg(\frac{\Delta}{2}+\epsilon_{2}+\eta\bigg), \qquad
 c\gamma_{2}=\frac{\Gamma+\Gamma_{2}+\Gamma_{3}}{4}+\xi
         -i\bigg(\frac{\Delta}{2}-\epsilon_{2}-\eta\bigg) \;,\\
&& \xi =  \frac{\sqrt{2}}{4}
  \left(\sqrt{\chi^{2}+4\Delta^{2}\Gamma^{\prime2}}-\chi\right)^{1/2}, \qquad\qquad\quad
\eta=\frac{\sqrt{2}}{4}
  \left(\sqrt{\chi^{2}+4\Delta^{2}\Gamma^{\prime2}}+\chi \right)^{1/2} \;,\\
&& \Gamma^{\prime} = \frac{\Gamma+\Gamma_{2}-\Gamma_{3}}{2}, \qquad\qquad\qquad\qquad\qquad\quad
\chi=\Delta^{2}+\Omega^{2}-\Gamma^{\prime2} \;.
\end{eqnarray}
\end{subequations}
\label{eq:lambda}
For the $\Lambda$-type 3LS and $N$-type 4LS cases, $C_{1,\,2}$ and $D_{1,\,2,\,3,\,4}$ take the same form
\begin{eqnarray}
 &&C_{1}^{(\Lambda,\,N)}(k_1,k_2) = \frac{\beta^{(\Lambda,\,N)}(k_1,k_2)-\alpha(k_1,k_2) \lambda_{2}}{\lambda_{1}-\lambda_{2}}, \qquad\qquad\qquad\qquad
 C_{2}^{(\Lambda,\,N)}(k_1,k_2)=\frac{-\beta^{(\Lambda,\,N)}(k_1,k_2)+\alpha(k_1,k_2) \lambda_{1}}{\lambda_{1}-\lambda_{2}},\nonumber \\
&& D_1^{(\Lambda,\,N)}(k_1,k_2,k_3)=\frac{\beta_{13}^{(\Lambda,\,N)}(k_1)-\alpha_{13}(k_1)\lambda_2}{\lambda_{1}-\lambda_{2}}C_{1}^{(\Lambda,\,N)}(k_2,k_3), \qquad
D_2^{(\Lambda,\,N)}(k_1,k_2,k_3)=\frac{-\beta_{24}^{(\Lambda,\,N)}(k_1)+\alpha_{24}(k_1)\lambda_1}{\lambda_{1}-\lambda_{2}}C_{2}^{(\Lambda,\,N)}(k_2,k_3),\nonumber \\
 &&D_3^{(\Lambda,\,N)}(k_1,k_2,k_3)=\frac{-\beta_{13}^{(\Lambda,\,N)}(k_1)+\alpha_{13}(k_1)\lambda_1}{\lambda_{1}-\lambda_{2}}C_{1}^{(\Lambda,\,N)}(k_2,k_3), \quad
 D_4^{(\Lambda,\,N)}(k_1,k_2,k_3)=\frac{\beta_{24}^{(\Lambda,\,N)}(k_1)-\alpha_{24}(k_1)\lambda_2}{\lambda_{1}-\lambda_{2}}C_{2}^{(\Lambda,\,N)}(k_2,k_3), \nonumber \\
&& \lambda_1=\frac{\Gamma+\Gamma_2-\Gamma_3}{4}+\xi
+i \bigg(\frac{\Delta}{2}+\eta \bigg),
\qquad\qquad\qquad\qquad\qquad
\lambda_2=\frac{\Gamma+\Gamma_2-\Gamma_3}{4}-\xi
+i \bigg(\frac{\Delta}{2}-\eta \bigg),
\end{eqnarray}
where the superscript $\Lambda$ stands for the 3LS and $N$ for the 4LS. $\alpha$'s and $\beta$'s in the above equation read
\begin{subequations}
\begin{eqnarray}
&&\alpha(k_1,k_2) =-\frac{(\overline{t}_{k_{1}}-1)(\overline{t}_{k_{2}}-1)}{2\pi}, \\
&&\beta(k_1,k_2)^{(\Lambda)}=\frac{\Gamma\Omega^{2}}{16\pi}\left[\frac{\overline{t}_{k_1}-1}{\rho_{k_2}}+\frac{\overline{t}_{k_2}-1}{\rho_{k_{1}}}\right], \qquad\quad
\beta(k_1,k_2)^{(N)}=\frac{\Gamma\Omega^{2}}{16\pi}\left[\frac{\overline{t}_{k_1}-\nu(k_1,k_2)}{\rho_{k_2}}+\frac{\overline{t}_{k_2}-\nu(k_1,k_2)}{\rho_{k_{1}}}\right], \\
&& \nu(k_1,k_2)=\frac{\epsilon_{4}-E-(i\Gamma_{4}-i\Gamma)/2}{\epsilon_{4}-E-(i\Gamma_{4}+i\Gamma)/2}, \qquad\qquad\qquad
\rho_{k}  = \left(ck-\epsilon_{2}+\Delta+\frac{i\Gamma_{3}}{2}\right)\left(ck-\epsilon_{2}+\frac{i\Gamma_{2}+i\Gamma}{2}\right)-\frac{\Omega^{2}}{4},
\end{eqnarray}
\end{subequations}
where $\overline{t}_{k}$ is given in Eq.\,(\ref{eq:SP_SE}b) in the main text. $\alpha_{13}$, $\alpha_{24}$, $\beta_{13}$ and $\beta_{24}$ are given by
\begin{subequations}
\begin{eqnarray}
 && \alpha_{13}(k)=\alpha_{24}(k)=-\frac{2(\overline{t}_{k}-1)}{\sqrt{2\pi}} \;, \\
&& \beta_{13}(k)^{(\Lambda)}=\frac{1}{\sqrt{2\pi}}\left[  \frac{\Gamma\Omega^2}{4\rho_{k}} -\left( \overline{t}_{k}-1 \right) \lambda_1 \right], \qquad\qquad
\beta_{13}(k)^{(N)}=\frac{1}{\sqrt{2\pi}}\left\{  \frac{\Gamma\Omega^2}{4\rho_{k}} -\left[ \overline{t}_{k}-\mu_1(k) \right] \lambda_1 \right\} \;, \\
&& \beta_{24}(k)^{(\Lambda)}=\frac{1}{\sqrt{2\pi}}\left[  \frac{\Gamma\Omega^2}{4\rho_{k}} -\left( \overline{t}_{k}-1 \right) \lambda_2 \right], \qquad\qquad
\beta_{24}(k)^{(N)}=\frac{1}{\sqrt{2\pi}}\left\{  \frac{\Gamma\Omega^2}{4\rho_{k}} -\left[ \overline{t}_{k}-\mu_2(k) \right] \lambda_2 \right\} \;, \\
&& \mu_{1,2}(k)=\frac{\epsilon_4-i\Gamma_4/2-ck+i\Gamma/2+ic\gamma_{1,2}}{\epsilon_4-i\Gamma_4/2-ck-i\Gamma/2+ic\gamma_{1,2}} \;.
\end{eqnarray}
\end{subequations}

\renewcommand{\theequation}{B\arabic{equation}}
\setcounter{equation}{0}  

\section*{Appendix B: Three-photon asymptotic output state}
In this Appendix, we present the asymptotic output state after scattering a three-photon right-going Fock state off a 3LS or 4LS. The form of the wave functions is
\begin{eqnarray}
&& |\psi^{(3)}\rangle = \int dk_1 dk_2 dk_3\frac{1}{\sqrt{3!}} \alpha(k_1)\alpha(k_2)\alpha(k_3) |\phi^{(3)}(k_1,k_2,k_3)\rangle, \nonumber \\
&& |\phi^{(3)}(k_1,k_2,k_3)\rangle = \int dx_1 dx_2 dx_3 \Big[\frac{1}{3!}ttt_{k_1,k_2,k_3}(x_1,x_2,x_3)a_R^{\dagger}(x_1)a_R^{\dagger}(x_2)a_R^{\dagger}(x_3)+ \frac{1}{2!}ttr_{k_1,k_2,k_3}(x_1,x_2,-x_3)a_R^{\dagger}(x_1)a_R^{\dagger}(x_2)a_L^{\dagger}(x_3)  \nonumber \\
&& \qquad +  \frac{1}{2!}trr_{k_1,k_2,k_3}(x_1,-x_2,-x_3)a_R^{\dagger}(x_1) a_L^{\dagger}(x_2)a_L^{\dagger}(x_3) + \frac{1}{3!}rrr_{k_1,k_2,k_3}(-x_1,-x_2,-x_3)a_L^{\dagger}(x_1) a_L^{\dagger}(x_2)a_L^{\dagger}(x_3)\Big] |\emptyset\rangle \;.
\end{eqnarray}
Here, $ttt_{k_1,k_2,k_3}(x_1,x_2,x_3)$, $ttr_{k_1,k_2,k_3}(x_1,x_2,x_3)$, $trr_{k_1,k_2,k_3}(x_1,x_2,x_3)$, and $rrr_{k_1,k_2,k_3}(x_1,x_2,x_3)$ are the terms representing three-photons being transmitted, two being transmitted and one reflected, one being transmitted and two reflected, and all three being reflected, respectively.
They take the following general form ($\alpha,\beta,\gamma=t\, \text{ or}\,r$)
\begin{eqnarray}
\label{eq:ttt}
&& \alpha\beta\gamma_{k_1,k_2,k_3}(x_1,x_2,x_3)=\sum_Q \alpha_{k_{Q_1}}\beta_{k_{Q_2}}\gamma_{k_{Q_3}}h_{k_{Q_1}}(x_1)h_{k_{Q_2}}(x_2)h_{k_{Q_3}}(x_3) + \frac{1}{4}\sum_Q\Big[\alpha_{k_{Q_1}} h_{k_{Q_1}}(x_1)B^{(2)}_{k_{Q_2},k_{Q_3}}(x_2,x_3) \nonumber \\
&&  \quad\quad\quad + \beta_{k_{Q_1}} h_{k_{Q_1}}(x_2)B^{(2)}_{k_{Q_2},k_{Q_3}}(x_1,x_3) + \gamma_{k_{Q_1}} h_{k_{Q_1}}(x_3)B^{(2)}_{k_{Q_2},k_{Q_3}}(x_1,x_2)\Big] + \frac{1}{8}\sum_{PQ} B^{(3)}_{k_{P_1},k_{P_2},k_{P_3}}(x_{Q_1},x_{Q_2},x_{Q_3}).
\end{eqnarray}
where $t_k$ and $r_k$ are the single-photon transmission and reflection probabilities given in Eq.\,(\ref{Output_SP}c), $B^{(2)}_{k_1,k_2}(x_1,x_2)$ is given in Eq.\,(\ref{eq:Bk1k2}), and $B^{(3)}_{k_1,k_2,k_3}(x_1,x_2,x_3)$ is given in Eq.\,(\ref{eq:g3}).
In Eq.\,(\ref{eq:ttt}), the first term comes from the process of three-photons passing by the atom as independent particles.
The second term corresponds to the process of one-photon passing through as an independent particle while the other two photons form a composite particle in a two-photon bound-state (with three possible combinations).
The third term originates from the three-photon bound-state process.

\renewcommand{\theequation}{C\arabic{equation}}
\setcounter{equation}{0}  
\section*{Appendix C: Second-order correlation function in the Schr\"{o}dinger picture}
In this Appendix, we demonstrate the equivalence between Eq.\,(\ref{eq:g2_H}) and Eq.\,(\ref{eq:g2_S}) in the main text. Typically, the second-order correlation function is defined in the
Heisenberg picture as,
\begin{equation}
g^{(2)}(x_{1},t_{1};x_{2},t_{2})=\frac{\langle\psi_0|\hat{a}^{\dagger}(x_{1},t_{1})\hat{a}^{\dagger}(x_{2},t_{2})\hat{a}(x_{2},t_{2})\hat{a}(x_{1},t_{1})|\psi_0\rangle}{\langle\psi_0|\hat{a}^{\dagger}(x_{1},t_{1})\hat{a}(x_{1},t_{1})|\psi_0\rangle\langle\psi_0|\hat{a}^{\dagger}(x_{2},t_{2})\hat{a}(x_{2},t_{2})|\psi_0\rangle}
\label{eq:g2_1}
\end{equation}
where $|\psi_0\rangle$ is the state in the Heisenberg picture,
or equivalently, the initial state in the Schr\"{o}dinger picture and
$\hat{a}^{\dagger}(x,t)$ is the operator in the Heisenberg picture.
$\hat{a}^{\dagger}(x,t)$ can be expressed in terms of the operator
in the Schr\"{o}dinger picture as
\begin{equation}
\hat{a}^{\dagger}(x,t)=e^{iHt/\hbar}\;\hat{a}^{\dagger}(x)\;e^{-iHt/\hbar}.
\label{eq:a_S_H}
\end{equation}
Taking $x_{1}=x_{2}=x$ in Eq.\,(\ref{eq:g2_1}), we obtain the two-time correlation function
\begin{equation}
g^{(2)}(x,t_{1};x,t_{2})=\frac{\langle\psi_0|\hat{a}^{\dagger}(x,t_{1})\hat{a}^{\dagger}(x,t_{2})\hat{a}(x,t_{2})\hat{a}(x,t_{1})|\psi_0\rangle}{\langle\psi_0|\hat{a}^{\dagger}(x,t_{1})\hat{a}(x,t_{1})|\psi_0\rangle\langle\psi_0|\hat{a}^{\dagger}(x,t_{2})\hat{a}(x,t_{2})|\psi_0\rangle}.
\label{eq:g2_2}
\end{equation}
If the field operator satisfies the following relation
\begin{equation}
\hat{a}^{\dagger}(x,t)=\hat{a}^{\dagger}(x-ct),
\label{eq:a_x_t}
\end{equation}
$g^{(2)}(x,t_{1};x,t_{2})$ is then the same as $g^{(2)}(x,t_{1};x^{\prime},t_{1})$
with $x^{\prime}=x-c(t_{2}-t_{1}).$ Using Eqs.\,(\ref{eq:a_S_H})
and (\ref{eq:a_x_t}), we can rewrite (\ref{eq:g2_2}) in
the Schr\"{o}dinger picture as
\begin{equation}
g^{(2)}(x,t_{1};x^{\prime},t_{1})=\frac{\langle\psi(t_{1})|\hat{a}^{\dagger}(x)\hat{a}^{\dagger}(x^{\prime})\hat{a}(x^{\prime})\hat{a}(x)|\psi(t_{1})\rangle}{\langle\psi(t_{1})|\hat{a}^{\dagger}(x)\hat{a}(x)|\psi(t_{1})\rangle\langle\psi(t_{1})|\hat{a}^{\dagger}(x^{\prime})\hat{a}(x^{\prime})|\psi(t_{1})\rangle},
\label{eq:g2_3}
\end{equation}
where $|\psi(t_{1})\rangle$ is the state at $t=t_1$ evolving from the initial state $|\psi_0\rangle$ under the Hamiltonian $H$.
Therefore, as long as Eq.\,(\ref{eq:a_x_t}) holds, the definition
of $g^{(2)}$ in the Heisenberg picture Eq.\,(\ref{eq:g2_2}) is
equivalent to Eq.\,(\ref{eq:g2_3}) defined in the Schr\"{o}dinger picture.
Physically, this means that
measuring the two-time correlation at the same spatial position is
equivalent to measuring the spatial correlation at the same time for
a non-dispersive field.

In our problem, it is straightforward to show that Eq.\,(\ref{eq:a_x_t}) is satisfied by the right-going field. With the Hamiltonian defined in Eq.\,(\ref{eq:Hamiltonian}) in the main text, the equation of motion for the right-going field in the 4LS case is
\begin{equation}
\left(\frac{\partial}{\partial x}+\frac{1}{c}\frac{\partial}{\partial t}\right)\hat{a}_{R}^{\dagger}(x,t)=\frac{iV}{c}\left[S_{12}^{+}(t)+S_{34}^{+}(t)\right]\delta(x).
\label{eq:EOM}
\end{equation}
Formally, the above equation can be integrated to yield
\begin{equation}
\hat{a}_{R}^{\dagger}(x,t)=\hat{a}_{R,free}^{\dagger}(x-ct)+\frac{iV}{c}\left[S_{12}^{+}(t-x/c)+S_{34}^{+}(t-x/c)\right]\theta(x).
\label{eq:aR}
\end{equation}
A similar expression can be obtained in the 3LS case.
Hence, Eq.\,(\ref{eq:a_x_t}) holds, and we use Eq.\,(\ref{eq:g2_3})
to evaluate the second-order correlation function of the transmitted field
with $|\psi(t_{1})\rangle$ being our final output state.

\end{widetext}

\vspace*{-2in}

\bibliography{3_4LS}
\end{document}